\documentclass[12pt]{article}
\usepackage{epsf,epsfig}

\newcommand{\be}{\begin{equation}}
\newcommand{\ee}{\end{equation}}
\newcommand{\ber}{\begin{eqnarray}}
\newcommand{\eer}{\end{eqnarray}}
\newcommand{\gsim}{\, \raisebox{-0.8ex}{$\stackrel{\textstyle >}{\sim}$ }}
\newcommand{\lsim}{\, \, \raisebox{-0.8ex}{$\stackrel{\textstyle <}{\sim}$ }}

\def\fm3{fm$^{-3}$}

\begin{document}
\baselineskip 16pt plus 2pt minus 2pt
\bibliographystyle{prsty2}

\begin{titlepage}

\par
\topmargin=-1cm      
{ \small

\noindent{NT@UW03-03} \hfill{INT-PUB 03-03}\\
}

\vspace{20.0pt}

\begin{centering}

{\large \bf Higher-order calculations of electron-deuteron scattering
in nuclear effective theory}

\vspace{40.0pt}
{\bf Daniel~R.~Phillips}\\
\vspace{20.0pt}
{\sl Department of Physics and Astronomy, Ohio
      University, Athens, OH 45701}\\
\vspace{2.0pt}
{\sl Department of Physics, University of Washington, Box 351560,
Seattle, WA 98195}\\
\vspace{2.0pt}
{\sl Institute for Nuclear Theory, University of Washington, 
Box 351550, Seattle, WA 98195}\\
\vspace{6.0pt}
{\tt Email: phillips@phy.ohiou.edu }\\

\vspace{15.0pt}
\end{centering}
\vspace{20.0pt}

\begin{abstract}
\noindent 
Motivated by recent advances in the application of effective field
theory techniques to light nuclei we revisit the problem of
electron-deuteron scattering in these approaches. By sidestepping
problems with the description of electron-nucleon scattering data in
effective field theories, we show that the effective theory expansion
for deuteron physics converges well over a wide range of momentum
transfers. The resultant description of the physics of the two-nucleon
system is good up to virtual photon momenta of order 700 MeV. 
\end{abstract}

\vspace*{10pt}
\begin{center}
\end{center}
\vfill

\end{titlepage}


\par
\topmargin=-1cm      

\vspace{20.0pt}

\section{Introduction}

Electron scattering from nuclei has a long and rich history. In
impulse approximation the charge form factor probed in such
experiments is the Fourier transform of the nuclear charge
distribution, and so these measurements have often been regarded as
independent tests of models of nuclear structure~\cite{deFW66,HS81}. In
particular, the structure of nuclei with $A \leq 10$ can now be
calculated {\it ab initio} from a given two- (and three-)nucleon
interaction~\cite{Pi02}. Calculations of electromagnetic form factors of these
nuclei then reveal agreement with experimental data that is, in general,
very good~\cite{WS98,CS98}

Here we focus on the simplest non-trivial nucleus: deuterium. Elastic
scattering of unpolarized electrons from deuterium results in an
$O(\alpha^2)$ differential cross-section:
\begin{equation}
\frac{d \sigma}{d \Omega}=\frac{d \sigma}{d \Omega}_{\rm Mott} \left[A(Q^2) + B(Q^2) \tan^2\left(\frac{\theta_e}{2}\right)\right],
\end{equation}
where $\theta_e$ is the electron scattering angle in the
centre-of-mass frame of the collision, $q^2=(p_e'-p_e)^2 \equiv -Q^2$ is the
(negative) virtuality of the (single) photon exchanged between the
electron and the nucleus, and $\frac{d \sigma}{d
\Omega}_{\rm Mott}$ is the Mott cross-section for electromagnetic
scattering from a point particle of charge $|e|$ and mass $M_d$. 

Deuterium is a spin-one nucleus and so has three independent form
factors. These are usually denoted by $G_C$, $G_Q$, and $G_M$. They
are related to Breit-frame matrix elements of the deuteron
electromagnetic current, $J_\mu$, through: 
\begin{eqnarray}
G_C&=&\frac{1}{3 |e|} \left(\left \langle 1\left|J^0\right|1 \right \rangle + 
\left \langle 0\left|J^0\right|0 \right \rangle + \left \langle -1\left|J^0\right|-1 \right \rangle
\right),\label{eq:GC}\\ 
G_Q&=&\frac{1}{2 |e|\eta M_d^2} 
\left(\left \langle 0\left|J^0\right|0 \right \rangle
- \left \langle 1\left|J^0\right|1 \right \rangle\right)\\ 
G_M&=&-\frac{1}{\sqrt{2 \eta} |e|} 
\left \langle1\left|J^+\right|0\right \rangle 
\label{eq:GM}
\end{eqnarray}
where we have labeled the deuteron states by the projection of the
deuteron spin along the direction of the three-vector ${\bf p}_e'
-{\bf p}_e$, and $\eta \equiv Q^2/(4 M_d^2)$. When defined in this
way these charge, quadrupole and magnetic form factors have the 
normalizations:
\begin{eqnarray}
G_C(0)&=&1,\\
G_Q(0)&=&Q_d,\\
G_M(0)&=&\mu_d \frac{M_d}{M};
\end{eqnarray}
where $Q_d=0.286~{\rm fm}^2$~\cite{BC79} is the deuteron quadrupole
moment, and $\mu_d=0.85741$~\cite{Li65} is the deuteron magnetic
moment in units of nuclear magnetons.

The experimental quantities $A$ and $B$ can then be
computed from theoretical models of deuterium, since
\begin{eqnarray}
A&=&G_C^2 + \frac{2}{3} \eta G_M^2 + \frac{8}{9} \eta^2  M_d^4 G_Q^2,\\
B&=&\frac{4}{3} \eta (1 + \eta) G_M^2.
\end{eqnarray}
However, it was not until the development of experiments
with polarized deuterium targets that it became possible
to unambiguously extract both $G_C$ and $G_Q$ from
electron-deuteron scattering data. The tensor-polarization
observable, $T_{20}$, is related to the ratios
\begin{eqnarray}
x &=& \frac{2}{3} \, \eta \, \frac{G_Q}{G_C},\\
y&=&\frac{2}{3} \, \eta \left(\frac{G_M}{G_C}\right)^2 
\left[\frac{1}{2} + (1 + \eta) \tan^2 \left(\frac{\theta_e}{2}\right) \right];
\end{eqnarray}
by
\begin{equation}
T_{20}=\sqrt{2} \frac{x(x + 2) + y/2}{1 + 2 (x^2 + y)},
\end{equation}
and so a measurement of $T_{20}$, together with measurements of $A$
and $B$ allows an extraction of $G_C$ and $G_Q$, and hence a complete
test of our theoretical understanding of deuteron
structure. Experiments over the last dozen years at
Bates~\cite{Sc84,Ga94}, Novosibirsk~\cite{Dm85,Gi90},
NIKHEF~\cite{Bo91,Fe96,Bo99}, and Jefferson Laboratory~\cite{Ab00A} have
measured $T_{20}$ in electron-deuteron scattering, and so facilitated
experimental determinations of the full set of deuteron structure
functions over a kinematic range between $Q=0$ and $Q=1.5$
GeV~\cite{Ab00B,Si01}. Modern nucleon-nucleon potentials, when
combined with models for two-body contributions to the deuteron
current, do a good job of reproducing this data (see,
e.g.~\cite{Wi95,Ar99,SP02}).  For a thorough status report on the
subject of electron-deuteron scattering we refer to three recent
reviews which discuss the subject~\cite{Si01,vOG01,GG01}.

In this paper we wish to address electron-deuteron scattering data in
the framework of effective theories of deuteron dynamics. This
approach (for recent reviews see Refs.~\cite{Be00,BvK02}) is based on
the use of a chiral expansion for the physics of the two-nucleon
system. Ultimately it shares many features with the more
``traditional'', and very successful, potential models. However, as
first suggested by Weinberg~\cite{We90,We91,We92}, this ``nuclear
effective theory'' is based on a systematic chiral and momentum
expansion for the the kernels of processes in the $NN$ system. Thus,
for electron-deuteron scattering we expand the deuteron current
$J_\mu$ in operators which are ordered according to their chiral
dimension, viz.:
\begin{equation}
J_\mu=e \sum_{i=1}^\infty c_i \frac{1}{\Lambda^{i-1}}{\cal O}_\mu^{(i)},
\label{eq:sum}
\end{equation}
where the operator ${\cal O}_\mu^{(i)}$ contains $i-1$ powers of the small
parameters $p$ (the momentum of the nucleons inside deuterium),
$m_\pi$, and $Q$. The numbers $c_i$ are, {\it a priori}, assumed to be
of order 1, and $\Lambda$ is the scale of chiral symmetry breaking:
$\Lambda \sim 4 \pi f_\pi, m_\rho, M$.  Since the expectation value of
$p$ and the value of $m_\pi$ are both much smaller than $\Lambda$
it follows that, provided $Q < \Lambda$, and the $c$s really are of
order one, this expansion should converge well. The expansion
parameter $(p,Q,m_\pi)/\Lambda$ is denoted here by $P$.

The operators ${\cal O}_\mu^{(i)}$ and the coefficients $c_i$ are
constructed according to the well-established counting rules and
Lagrangian of chiral perturbation theory~\cite{Be95}. Here we present
results for $G_C$ and $G_Q$ up to order $eP^3$, and results for $G_M$
up to $O(eP^2)$. We go beyond the recent calculation of
Ref.~\cite{WM01}, which computed all three form
factors only up to $O(e P^2)$. We also demonstrate that, provided
single-nucleon structure effects are correctly included in the
calculation, the nuclear effective theory is, in fact, much more
accurate than the results of Ref.~\cite{WM01} might lead one to
believe. Indeed, ultimately it describes all of the extant
experimental data on $G_C$ and $G_Q$ out to momentum transfers of
order 700 MeV.

This is done as follows. In Section~\ref{sec-kernel} we sketch the
derivation of $J_\mu$ from the counting rules of chiral perturbation
theory, and give results for the current at leading order, $O(e)$,
next-to-leading order $O(e P^2)$, and next-to-next-to-leading order,
$O(e P^3)$.  In Section~\ref{sec-wfs} we will discuss the wave
functions used in our calculation, and outline some of the issues
associated with the desire for consistency between the deuteron
current and the deuteron wave functions. In Section~\ref{sec-results}
we will present our results for $G_C$, $G_Q$, and $G_M$, as well as
results for the deuteron's static properties $\mu_d$, $Q_d$, and the
deuteron charge radius. We conclude in Section~\ref{sec-conclusion}.

\section{The deuteron current}

\label{sec-kernel}

The heavy-baryon chiral perturbation theory (HB$\chi$PT) Lagrangian is
organized according to the powers of $P$ which appear in the classical
Lagrange density. The  pieces of the leading-order ($O(P)$)
heavy-baryon Lagrangian relevant to the computation to be
presented here are:
\begin{equation}
{\cal L}_{\pi N}^{(1)}=N^\dagger (i v \cdot D) N + g_A N^\dagger u \cdot S N,
\label{eq:LpiN1}
\end{equation}
with:
\begin{eqnarray}
D_\mu&=&\partial_\mu - \frac{ie}{2} (1 + \tau_3) A_\mu + \ldots\\ 
u_\mu&=&i u^\dagger \partial_\mu U u^\dagger,
\end{eqnarray}
and $v$ chosen to be $v=(1,0,0,0)$, so that:
\begin{equation}
S=(0,{\bf S}); \quad {\bf S}= \frac{\bf \sigma}{2},
\end{equation}
We also choose the pion interpolating field such that:
\begin{equation}
u^2=U=\exp\left(\frac{i \vec{\tau} \cdot \vec{\pi}}{f_\pi}\right).
\end{equation}
$A_\mu$ is the photon field. Note that we have omitted some terms that are
of higher-order in the pion field than we need for our calculation.

The part of ${\cal L}_{\gamma N}^{(2)}$ relevant for our calculation
is the photon-nucleon piece. There we focus on the vertices,
suppressed by order $p,Q/M$, that govern the coupling of E1 and M1
photons to the nucleon~\cite{Br95}:
\begin{equation}
{\cal L}_{\gamma N}^{(2)}=N^\dagger \frac{1}{2 M}
\left[(v \cdot D)^2 - D \cdot D\right] N
- \frac{ie}{4M} N^\dagger [S_\mu,S_\nu] \left[(1 + \kappa_v) \tau_3
 + (1 + \kappa_s)\right] F^{\mu \nu} N,
\end{equation}
with $F_{\mu \nu}$ the electromagnetic field strength tensor.
$\kappa_s$ and $\kappa_v$ are the isoscalar and isovector parts of the
anomalous magnetic moment of the nucleon. These are known
experimentally, and have the values $-0.12$ and $3.90$, respectively.

There is also an important term whose coefficient is entirely
determined by reparameterization invariance. It occurs after the
Foldy-Wouthysen transformation is used to eliminate the
lower-component of the heavy-baryon field~\cite{Br95}:
\begin{equation}
{\cal L}_{FW}^{(2)}=-N^\dagger \frac{i g_A}{2M}
\left\{ S \cdot D, v \cdot u\right\} N.
\end{equation}
Employing the definitions above, then reorganizing the result by
eliminating total derivatives and using the nucleon equation of motion,
leads to the piece relevant for our study:
\begin{equation}
{\cal L}_{\pi \gamma N}^{(2)}=\frac{e g_A}{2 M f_\pi} N^\dagger
\tau^a \pi^a ((S \cdot \partial) v \cdot A) N.
\end{equation}

The first occurrence of the finite electric radius of the isoscalar
nucleon occurs in chiral perturbation theory as a coefficient in the
Lagrangian ${\cal L}_{\gamma N}^{(3)}$. Similarly, the magnetic radius
of the nucleon appears as a coefficient in ${\cal L}_{\gamma
N}^{(4)}$. In both of these Lagrangians one also encounters terms
arising from relativistic corrections to the single-nucleon
four-current. The coefficients of these structures are determined by
reparameterization invariance, and can be found by taking the
relativistic current operator and using the standard procedure for
generating the non-relativistic one-body current operator as an
expansion in powers of $p/M$ and $Q/M$ (see, for instance
\cite{CS98,AA96}).

Finally, in ${\cal L}^{(4)}$ we encounter a two-nucleon
operator representing a magnetic photon coupling to the $NN$
system~\cite{WM01,Sc99}:
\begin{equation}
{{\cal L}_{\gamma NN}^{(2)}}_{\cal M}=
-i e L_2 (N^\dagger [S_\mu,S_\nu] F^{\mu \nu}
N)(N^\dagger N).
\label{eq:LMNN}
\end{equation} 
This short-distance two-body current will modify the magnetic moment
of deuterium.  

Similarly, in ${\cal L}_{\gamma NN}^{(3)}$ 
there is an operator which represents a
quadrupole (E2) photon coupling to the $NN$ system~\cite{Ch99} and so
modifies the deuteron quadrupole moment. At the same order there is
also an operator  which modifies the deuteron charge radius~\cite{Ch99}. 

The vertices derived from the Lagrangians
(\ref{eq:LpiN1})--(\ref{eq:LMNN}) are then used to draw all possible
Feynman diagrams contributing to the process $\gamma^* NN
\rightarrow NN$. A particular Feynman diagram then leads to an
operator appearing in the sum (\ref{eq:sum}). The power of $P$ that
this operator possesses is defined by considering all parts of the
amputated Feynman diagram representing it, and multiplying together
the ``$P$-scaling factors'' of these separate pieces. These factors
are defined as follows:
\begin{itemize}
\item A vertex from $L_{\pi N}^{(n)}$ contributes a factor of $P^n$.

\item A vertex involving a photon from $L_{\gamma N}^{(n)}$,
$L_{\gamma \pi N}^{(n)}$, or $L_{\gamma NN}^{(n)}$ contributes a factor of
$P^{n-1}$~\cite{factore}.

\item Each pion propagator contributes a factor of $P^{-2}$.

\item Each nucleon propagator contributes a factor of $P^{-1}$. 

\item A two-body graph has an additional factor of $P^3$.

\item Each loop contributes a factor of $P^4$.
\end{itemize}

\begin{figure}[htbp]
\vspace{0.5cm}
\centerline{\epsfig{figure=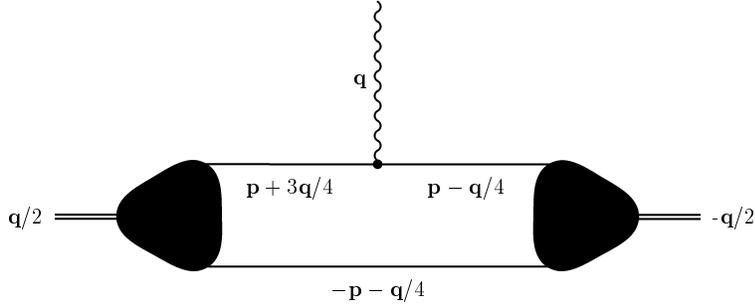,width=10.0cm}}
\vspace*{-1.0cm}
\caption{Three momenta of the deuteron, photon, and nucleons in the
Breit frame for a generic one-body contribution to $J_\mu$. This frame
is chosen because in it the photon is purely space-like: $q=(0,{\bf
q})$. Time runs from right to left.}
\label{fig-Breit}
\end{figure}

We now discuss the charge and current operators in turn.  Such a
decomposition is, of course, not Lorentz invariant, so here we make
this specification in the Breit frame. where the three-momentum of the
deuteron and the nucleons is as shown in Fig.~\ref{fig-Breit}.

\begin{figure}[htbp]
\vspace{0.2cm}
\hspace{-0.25in}
\centerline{\epsfig{figure=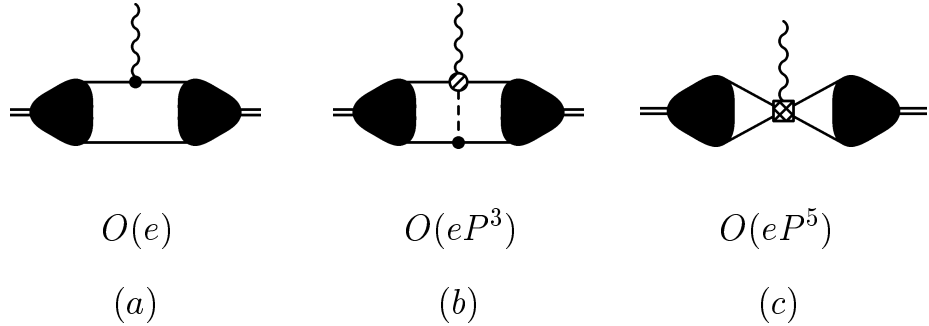,width=11.0cm}}
\vspace{-0.1cm}
\caption{Diagrams representing the leading contribution
to the deuteron charge operator [(a)], the leading two-body contribution 
to $J_0$ [(b)], and the dominant short-distance piece [(c)].
Solid circles are vertices from ${\cal L}_{\pi N}^{(1)}$, and the shaded
circle is the vertex from ${\cal L}_{\gamma \pi N}^{(2)}$. The hatched
square is a four-nucleon vertex from ${\cal L}_{\gamma NN}^{(3)}$.}
\label{fig-twobodycharge}
\end{figure}

\paragraph {Deuteron charge}
The vertex from ${\cal L}_{\pi N}^{(1)}$ which represents an $A_0$ photon
coupling to the nucleon gives the leading-order (LO) contribution to $J_0$:
\begin{equation}
J_0^{(0)}=|e|.
\end{equation}
This is depicted in Fig.~\ref{fig-twobodycharge}(a).

The most important correction to $J_0$ arises from the insertions in
${\cal L}_{\pi N}^{(3)}$ which generate the nucleon's isoscalar charge
radius. This gives a result for $J_0$ through $O(eP^2)$:
\begin{equation}
{J_0}_{\rm structure}^{(2)}=|e| \left(1 - \frac{1}{6} \langle r_{Es}^2 \rangle Q^2\right),
\label{eq:structure}
\end{equation}
where $\langle r_{Es}^2 \rangle$ is the isoscalar charge radius of the nucleon,
for which we adopt the value:
\begin{equation}
\langle r_{Es}^2 \rangle=(0.777~{\rm fm})^2.
\end{equation}
(Note: $Q^2={\bf q}^2$ holds in the Breit frame.)

Also present at this order are relativistic corrections to the
single-nucleon charge operator. To generate the ``intrinsic'' current
operator which can be inserted between deuteron wave functions
calculated in the two-nucleon center-of-mass frame we employ the
formalism of Adam and Arenh\"ovel, as described in Ref.~\cite{AA96}. The
relativistic corrections then fall into two categories: corrections
coming from the expansion of the relativistic single-nucleon current
in powers of $p/M$, and corrections due to the necessity of boosting
the deuteron wave function from the frame where ${\bf P}={\bf 0}$ to
the frame where ${\bf P}=\pm{\bf q}/2$.

When the calculation is organized in this way the dominant
``relativistic effect'' for momentum transfers of order 500 MeV is a
shift in the length of ${\bf q}$. This ``length contraction'' accounts
for a portion of the boost of the deuteron wave function (for details see
Refs.~\cite{SP02,AA96}). The net result is that whereas the leading-order form
factor $G_C$ can be represented as:
\begin{equation}
G_C^{(0)}=|e| \int \frac{d^3 p}{(2 \pi)^3} \psi^*\left({\bf p} + \frac{\bf
q}{2}\right) \psi({\bf p}),
\end{equation}
with $\psi$ the deuteron wave function, at $O(e P^2)$ the
expression is:
\begin{equation}
{G_C}_{\rm boost}^{(2)}=|e| \int \frac{d^3 p}{(2 \pi)^3}
\psi^*\left({\bf p} + \frac{\bf q}{2 \sqrt{1 + \eta}}\right) \psi({\bf
p}),
\end{equation}
where $\eta=Q^2/(4 M_d^2)$ was defined above. Here we have not
reproduced the terms which scale as $p/M$, and we have not included the
terms from Eq.~(\ref{eq:structure}). The sole effect written is the one arising
from the boost of the deuteron wave function, although all effects occurring
at $O(e P^2)$ are included in our computation. 

This completes the discussion of mechanisms contributing at $O(e
P^2)$, or next-to-leading order. At $O(e
P^3)$---next-to-next-to-leading order---the Lagrangian
(\ref{eq:LpiN1}) generates a tree-level two-body graph with an
isoscalar structure, as shown in Fig.~\ref{fig-twobodycharge}(b).
This two-body contribution to $J^0$ was derived by Riska in
Ref.~\cite{Ri84}, using an argument based on matching to relativistic
Born graphs for pion electroproduction. Here it occurs in HB$\chi$PT
as a natural consequence of the Foldy-Wouthysen transformation which
generates the relevant term in ${\cal L}^{(2)}$. Importantly, the
nuclear effective theory also has the ability to organize the
contribution of two-body contributions, such as this, relative to the
contribution of one-body mechanisms.

Straightforward application of the Feynman rules for the relevant
pieces of the HB$\chi$PT Lagrangian gives the result for this
piece of the deuteron current:
\begin{equation}
\langle {\bf p}'|J_0^{(3)}({\bf q})|{\bf p} \rangle=
\tau_1^a \tau_2^a \, \, \frac{|e| g_A^2}{8 f_\pi^2 M}  \, \,  
\left[\frac{\sigma_1 \cdot {\bf q} \, \sigma_{2} \cdot 
({\bf p} - {\bf p}' + {\bf q}/2)}
{m_\pi^2 + ({\bf p} - {\bf p}' + {\bf q}/2)^2} + (1 \leftrightarrow 2)\right],
\label{eq:J02B}
\end{equation}
where ${\bf p}$ and ${\bf p}'$ are the (Breit-frame) relative momenta
of the two nucleons in the initial and final-state
respectively~\cite{unitequivnote}.

The short-distance two-body currents that contribute to $\langle r_d^2
\rangle$ and $Q_d$ are depicted in Fig.~\ref{fig-twobodycharge}(c).
They do not give a contribution until $O(e P^5)$. This suggests that
the charge operator is not particularly sensitive to short-distance
physics, since two-body effects of range $1/\Lambda$ are suppressed by
five powers of $P$ relative to the LO result.

\paragraph{Deuteron three-current} The counting for the isoscalar 
three-vector current ${\bf J}$ was already considered in detail by
Park and collaborators~\cite{Pa95}. ${\bf J}$ begins at $O(eP)$. There the
operator derived from ${\cal L}_{\gamma N}^{(2)}$ is:
\begin{equation}
{\bf J}^{(1)}=|e| ({\bf p} + {\bf q}/2)/M + i \mu_S {\bf \sigma}
\times {\bf q},
\label{eq:J1}
\end{equation}
where ${\bf p} - {\bf q}/4$ is the momentum of the struck nucleon, as
shown in Fig.~\ref{fig-Breit}, and $\mu_S$ is the isoscalar magnetic
moment of the nucleon, whose value we take to be $\mu_S= 0.88 |e|/(2M)$.

\begin{figure}[htbp]
\vspace{0.2cm}
\hspace{-0.3in}
\centerline{\epsfig{figure=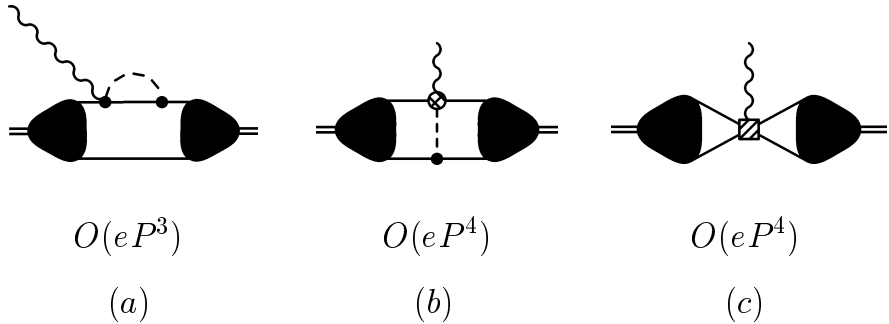,width=11.0cm}}
\vspace{-0.1cm}
\caption{An $O(eP^3)$ loop diagram which ultimately does not 
generate $q^2$-dependence in the nucleon isoscalar form factor [(a)],
and two contributions of order $O(eP^4)$ [(b) and (c)]. The hatched
circle is a vertex from ${\cal L}_{\pi \gamma N}^{(3)}$, while the shaded
square is the vertex from ${{\cal L}_{\gamma NN}^{(2)}}_{\cal M}$.}
\label{fig-magnetic}
\end{figure}

As in the case of $J_0$, there are finite-size and relativistic
corrections to Eq.~(\ref{eq:J1}) which are suppressed by two powers of
$P^2$. Thus, in this case they enter at $O(e P^3)$, and represent the
NLO corrections to $G_M$. Loop graphs of the type depicted in
Fig.~\ref{fig-magnetic}(a) also enter at this order. However, it can
be shown that the only effect of these loops on the isoscalar $NN$
current is to renormalize the magnetic moment of the nucleon: their
isoscalar part does not have any $q^2$ dependence (an analogous
argument is given for real photons in Ref.~\cite{Pa95}).

At $O(e P^4)$ [NNLO] two kinds of magnetic two-body current enter the
calculation. Park {\it et al.} have pointed out that when magnetic
photons interact with deuterium there is a single-nucleon $\gamma \pi$
contact term in ${\cal L}_{\pi \gamma N}^{(3)}$~\cite{Pa95}. The coefficient
of this portion of the chiral Lagrangian was fixed in Ref.~\cite{Pa95}
using the KSRF relation and a resonance-saturation hypothesis.
Alternatively, this coefficient could also be fixed by comparison to
data---at least in principle. In either case this $\gamma \pi NN$
vertex generates a pion-range two-body current with a coefficient that
is undetermined {\it a priori}, as shown in
Fig.~\ref{fig-magnetic}(b).

Meanwhile, a number of authors~\cite{WM01,Sc99,Ch99,Pa95,Pa00}, have
pointed out that the short-distance two-body operator from the
Lagrangian in Eq.~(\ref{eq:LMNN}) contributes to ${\bf J}$ at
$O(eP^4)$. It generates a ``short-range'' exchange-current
contribution to $G_M$ (see Fig.~\ref{fig-magnetic}(c)). Since this is
only suppressed by $P^3$ relative to the leading contributions to
$G_M$ we would expect $G_M$ to be markedly more sensitive to details
of the short-distance physics than $G_C$. Given the presence of two
undetermined parameters at NNLO in ${\bf J}$ we will only examine the
leading and next-to-leading order predictions of the nuclear effective
theory for $G_M$.

\section{Deuteron wave functions}

\label{sec-wfs}

In order to define the computation completely it remains only to
specify the deuteron wave functions which will be used for the
evaluation of the matrix elements in
Eqs.~(\ref{eq:GC})--(\ref{eq:GM}). Here we will use four different
kinds of wave function:
\begin{enumerate}
\item A ``strict'' chiral perturbation theory wave function, as
derived in Ref.~\cite{Ep99}. We generally employ the NLO wave
function, with the cutoff chosen to be $\Lambda=600$ MeV. We also use
Epelbaum et al.'s NLO wave function with $\Lambda=540$ MeV for
comparison.

\item The N$^2$LO wave function of Ref.~\cite{EM01}. In this
calculation a specific choice of cutoff is made, which allows for
better accuracy in fitting $NN$ phase shifts. Certain relativistic
corrections to the $NN$ potential are also included.

\item The wave functions derived in Ref.~\cite{PC99} by ``integrating
in'' the one-pion exchange potential (OPEP) to a given radius
$R$. These should be regarded as very simplistic potential models for
deuterium. They are, however, designed to produce the correct values
for the important deuteron properties $A_S$, $A_D$, and $B$, as well
as to include the standard non-relativistic OPEP (with the ``modern''
coupling constant).

\item The deuteron wave function obtained using the Nijm93 meson-theoretic
potential~\cite{St93}. 
\end{enumerate}

There is an important question of consistency with the current for all
of these wave functions. In particular, it is well known that the
charge contribution (\ref{eq:J02B}) is associated with so-called
``relativistic corrections'' to the one-pion-exchange potential.  If
$J_0^{(3)}$ and the terms in OPEP suppressed by $p^2/M^2$ relative to
the leading behaviour are derived within a consistent framework then
the results for deuteron form factors should be unitarily
equivalent~\cite{Fr80,Ad93,Fo99}. In fact, the authors of
Ref.~\cite{Ep99} did not consider ``relativistic corrections'' to
one-pion exchange.  They counted $M \sim \Lambda_\chi^2$ and so
regarded the pieces of OPEP of relative order $p^2/M^2$ as being down
by $P^4$ compared with the leading piece of the chiral $NN$
potential. Indeed, none of the wave functions listed under numbers
one, three, and four above include any ``relativistic corrections'' to
one-pion exchange. Clearly a fully-consistent treatment of the
deuteron current and $NN$ potential in $\chi$PT which incorporates
what has been learned about unitary equivalence~\cite{Fr80,Ad93,Fo99}
is necessary if a definitive result is to be established for
electron-deuteron scattering in the nuclear effective theory.

Here our goal is less ambitious. We take wave functions presently on
the market and use the expansion for the deuteron current discussed in
Sec.~\ref{sec-kernel} to generate results for $G_C$, $G_Q$, and
$G_M$. The error resulting from inconsistencies in this procedure can
be assessed by comparing the results obtained with the wave functions
of Refs.~\cite{Ep99,PC99,St93} to those found using the ``Idaho'' wave
function of Ref.~\cite{EM01}. Of the wave functions used here, only
the ``Idaho'' wave function includes the effect of relativistic
corrections to one-pion exchange of the type associated by unitary
equivalence with the two-body charge contribution (\ref{eq:J02B}).

\section{Results}

\label{sec-results}

\paragraph{Strict chiral expansion}

\begin{figure}[htbp]
\centerline{\epsfig{figure=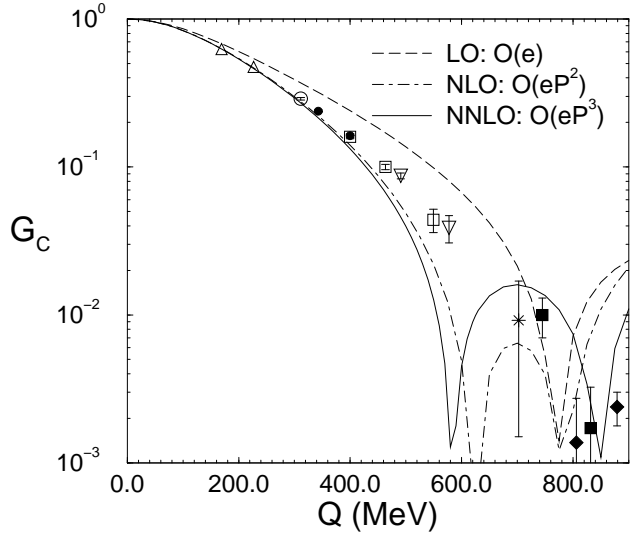,width=8.0cm}}
\caption{$G_C$ as calculated in the strict $\chi$PT expansion for
$J_0$ at leading, next-to-leading, and next-to-next-to-leading
order, plotted against $|{\bf q}|$. The experimental data
is taken from the extraction of Ref.~\cite{Ab00B}: upward triangles
represent data from the $T_{20}$ measurement of Ref.~\cite{Dm85}, open
circle \cite{Fe96}, solid circle \cite{Sc84}, open squares
\cite{Bo99}, downward triangles \cite{Gi90}, star \cite{Bo91}, solid
squares \cite{Ga94}, solid diamonds \cite{Ab00A}.}
\label{fig-GCbreakdown}
\end{figure}

The results of the leading-order (LO), next-to-leading order (NLO),
and next-to-next-to-leading order calculations for $G_C$, using the
NLO $\chi$PT wave function of Ref.~\cite{Ep99} with $\Lambda=600$ MeV,
are displayed in Fig.~\ref{fig-GCbreakdown}. Also shown there are data
from the compilation~\cite{Ab00B}.  The $\chi$PT expansion for $J_0$
appears to be converging for $q \leq 600$ MeV, but it is not
converging to the data. As was already observed in the NLO
calculations of Walzl and Mei\ss ner~\cite{WM01}, a strict
chiral expansion of $J^0$ does a poor job of describing 
data on $G_C$ for $Q^2 > 0.1~{\rm GeV}^2$.

The reason for this failure can be traced to the isoscalar nucleon
form factor obtained in $\chi$PT~\cite{Br92}. That form factor is:
\begin{equation}
G_E^N(Q^2)=1 - \frac{1}{6} \langle r_N^2 \rangle Q^2,
\end{equation}
and it describes electron-nucleon scattering data only up to
$Q^2 \sim 0.1~{\rm GeV}^2$. The inclusion of heavy mesons in the chiral
lagrangian remedies this situation somewhat~\cite{KM99}, but if we
insist on a strict $\chi$PT expansion---or even include explicit Delta
degrees of freedom in the theory~\cite{He98}---our description of
electron-deuteron data will be limited by $\chi$PT's difficulty in
describing single-nucleon isoscalar electromagnetic structure.

\paragraph{Factorization}
A solution to this problem is provided by the factorization of
$J_0$. Up to the order to which we work the deuteron charge operator
can be written as the product of a piece that describes the current
due to structureless nucleons and a nucleon-structure piece:
\begin{equation}
\langle {\bf p}'|J_0|{\bf p}\rangle=
(|e| \delta^{(3)}(p' - p - q/2)
+ \langle{\bf p}'|J_0^{(3)}({\bf q})|{\bf p} \rangle) G_E^N(Q^2) 
+ O(eP^4).
\end{equation}
(Relativistic corrections are not written here, but, at this order,
factorization is also valid for them.)

\begin{figure}[htbp]
\centerline{\epsfig{figure=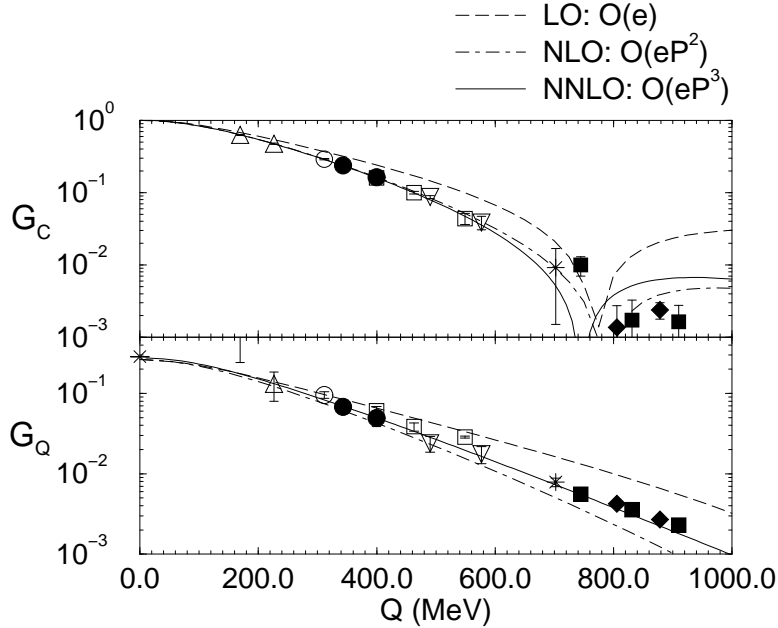,width=10.0cm}}
\caption{$G_C$ and $G_Q$ (in units of fm$^2$) calculated with nucleon
structure effects included via factorization (at NLO and NNLO, LO is
as before). The NLO $\chi$PT deuteron wave function with $\Lambda=600$
MeV was used.  Legend as in Fig.~\ref{fig-GCbreakdown}.}
\label{fig-GCGQ}
\end{figure}

Here we focus on the ability of nuclear effective theory to describe
deuteron structure, and so we choose to apply the chiral expansion to
the ratios of form factors:
\begin{equation}
\frac{G_C}{G_E^N} \quad \mbox{and} \quad \frac{G_Q}{G_E^N}.
\label{eq:ratios}
\end{equation}
To do this we compute the ratio $J_0/G_E^N$, i. e. the electric
response of a deuterium nucleus containing structureless nucleons.
Then, in order to compare with the compilation~\cite{Ab00B}, the
ratios (\ref{eq:ratios}) are multiplied by the parameterization of
$G_E^N$ found in Ref.~\cite{Me95}.  The results obtained by this
procedure are shown in Fig.~\ref{fig-GCGQ}. This time the expansion
not only converges, provided that $Q \lsim 700$ MeV, but also
reproduces data on both $G_C$ and $G_Q$ in this range of $Q$.

Expanding the quantities (\ref{eq:ratios}) in the effective theory
sidesteps $\chi$PT's problems in describing isoscalar nucleon
structure. We find that the chiral expansion for these ratios is in
good agreement with data. Since these are the type of quantities which
must be calculated in order to extract nucleon-structure
information from deuteron data the results shown in
Fig.~\ref{fig-GCGQ} are quite encouraging in this regard.

\begin{figure}[htbp]
\centerline{\epsfig{figure=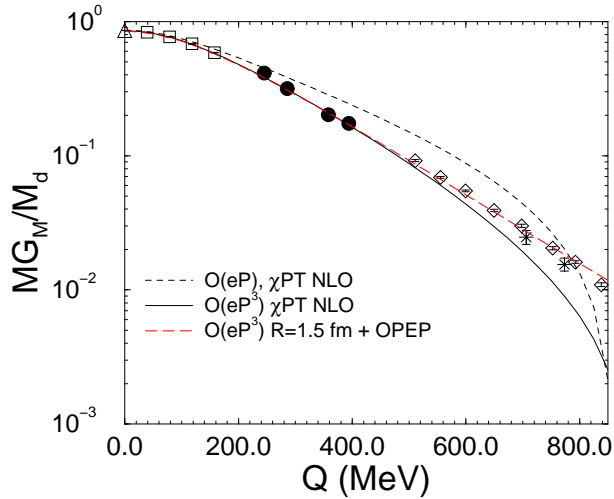,width=8.0cm}}
\caption{The deuteron magnetic form factor as calculated to LO with
the $\chi$PT NLO wave function (black short-dashed line) and NLO with
the $\chi$PT NLO wave function (solid black line) and the $R=1.5$ fm +
OPEP wave function (red long-dashed line). Factorization is used to
include nucleon structure in the NLO results. Experimental data from
deuteron magnetic moment, open triangle~\cite{Li65}; the
parameterization of Ref.~\cite{Si01}, open squares; and measurements
of $B(Q^2)$: solid circles~\cite{Si81}, open diamonds~\cite{Au85}, and
stars~\cite{Cr85}.}
\label{fig-GM}
\end{figure}

Turning to the magnetic form factor, factorization also holds there,
to the order to which we work, and so we compute the chiral expansion
for the ratio $J^+/G_M^N$. Since we only calculate $J^+$ to NLO it is
difficult to judge the convergence of the series, but the description
of the data is quite good over the range $Q \lsim 500$ MeV.

\paragraph{Estimating the size of short-distance effects}
In order to judge the sensitivity of this observable to short-distance
effects, in Fig.~\ref{fig-GM} we also show the result for $G_M$
obtained with a simple short-distance + OPEP wave
function~\cite{PC99}. This wave function and the $\chi$PT NLO wave
function differ only at distances $r \ll 1/m_\pi$, and so the
red-dashed line's agreement with data to $Q \sim 900$ MeV should be
regarded as fortuitous.  From an EFT point of view, the difference
between the red-dashed and solid curves in Fig.~\ref{fig-GM} is a
short-distance effect. Such effects enter at NNLO in this observable,
and so the sizable impact of physics at distances $r \sim 1/\Lambda$
on $G_M$ that is seen in Fig.~\ref{fig-GM} is not surprising.

\begin{figure}[htbp]
\centerline{\epsfig{figure=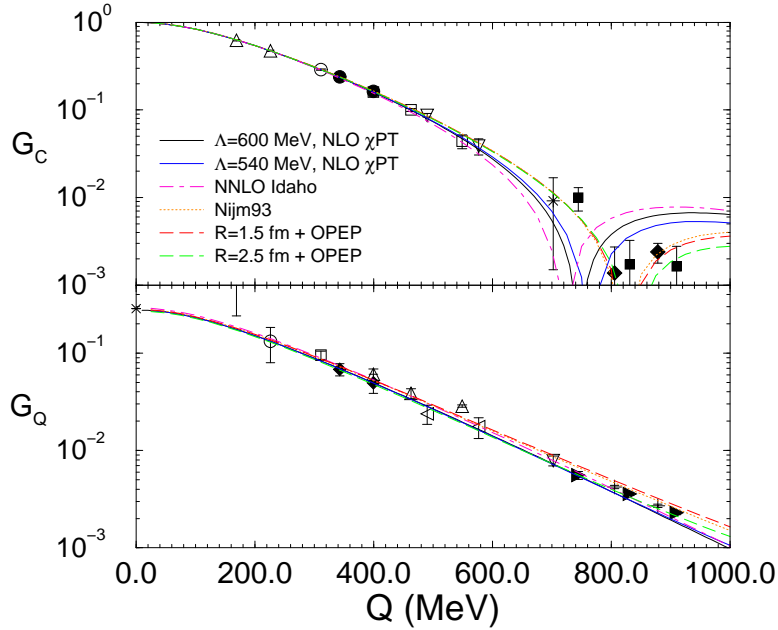,width=10.0cm}}
\caption{Results with different wave functions for $G_C$ and $G_Q$ (in
units of fm$^2$).  Solid black and blue lines are with wave functions
from Ref.~\cite{Ep99}, purple dot-dashed with that of \cite{EM01},
dotted \cite{St93}, and red and green long-dashed \cite{PC99}.}
\label{fig-GCGQcomp}
\end{figure}

In contrast, short-distance contributions to $G_C$ and $G_Q$ do not
occur until $O(eP^5)$. As with $G_M$, we can
estimate their impact by computing the form factors with different
deuteron wave functions---see Fig.~\ref{fig-GCGQcomp}.  The results
for $G_C$ and $G_Q$ are largely the same for $Q \lsim 600$ GeV. The
most noticeable difference occurs around the zero of $G_C$---a region
where sensitivity to details of deuteron physics is well-established.

Intriguingly, the band representing different assumptions about
short-distance physics is quite narrow out to values of
$Q \gsim 800$ MeV when $G_Q$ is considered. This suggests that the shape of
$G_Q$ is not strongly affected by short-distance physics, and
higher-order corrections to it may well be small. (A similar
conclusion was reached without the use of nuclear effective theory in
Ref.~\cite{SS01}.)

The curves of Figs.~\ref{fig-GCGQ}--\ref{fig-GCGQcomp} are not,
strictly speaking, predictions of $\chi$PT for $G_C$ and $G_Q$. In
terms of the chiral expansion for these form factors a particular
class of higher-order terms for electron-nucleon scattering have been
resummed: the class of terms responsible for reproducing the
``experimental'' $G_E^N$.  Nevertheless, the results of the procedure
we have adopted show that nuclear effective theory does a good
job of describing {\it deuteron} structure---and especially the deuteron
charge distribution---out to surprisingly high momentum transfers.

\paragraph{Deuteron static properties:}

As far as deuteron static properties are concerned it is irrelevant how
nucleon structure is included in the calculation. We have computed:
\begin{equation}
\langle r_d^2 \rangle \equiv -6 \left. \frac{d G_C}{d Q^2}\right|_{Q^2=0},
\end{equation}
when $G_C$ is calculated with structureless nucleons. The result for
$r_d \equiv \langle r_d^2 \rangle^{1/2}$ is shown in
Table~\ref{table-chiralstatic}, together with results for $\mu_d$ and
$Q_d$.  Once again, the convergence of the expansion is very good,
with the leading-order result capturing most of the physics of these
static properties.

\begin{table}[phtb]
\begin{center}
\vskip 0.6 true cm 
\begin{tabular}{||c||c|c|c||}
    \hline \hline
$J_\mu$ Order & $r_d$ (fm) & $\mu_d$ (n.m.) & $Q_d$ (fm$^2$) \\
\hline  \hline
LO    &  1.975     &     0.8591     &  0.2660  \\
NLO   &  1.984     &     0.8531     &  0.2641  \\
NNLO  &  1.987     &   Experiment   &  0.2764  \\ 
\hline \hline
\end{tabular}
\end{center}
\caption{Deuteron static properties computed with the NLO $\chi$PT
wave function ($\Lambda=600$ MeV) at LO, NLO, and NNLO. At NNLO
$\mu_d$ can be exactly reproduced by adjusting the coefficient $L_2$
in Eq.~(\ref{eq:LMNN}). (The numerical error in each quantity is $\pm
1$ in the last significant figure quoted.)}
\label{table-chiralstatic}
\end{table}

In order to assess the sensitivity of these quantities to
short-distance effects we have computed $r_d$, $\mu_d$, and $Q_d$ with
a variety of deuteron wave functions. The results are summarized in
Table~\ref{table-compstatic} and agreement with experimental data is
very good. The $\sim 0.5$\% discrepancy in $r_d$ is certainly consistent
with the expected size of the $P^4$ corrections omitted here, while
the $\sim 1$\% discrepancy in $\mu_d$ is perhaps less than one would
naively expect, given the that effects of relative order $P^3$ 
were not included in this computation of the deuteron's magnetic moment.

\begin{table}[phtb]
\begin{center}
\vskip 0.6 true cm 
\begin{tabular}{||l||c|c|c|c|c|c||}
\hline \hline
 & Experiment & Nijm 93 & $\chi$PT NLO & NNLO Idaho  & OPEP + short\\
& & & $\Lambda=600$ MeV & & $R=1.5$ fm\\
\hline  \hline
$A_S$ (fm$^{-1/2}$)&  0.8846(8) & 0.8842   & 0.869  & 0.885  &0.8845
\\ \hline
$A_D/A_S$          &  0.0256(4) & 0.0252   & 0.0248 & 0.0245 &0.0253
\\ \hline
$B$ (MeV)          & 2.224575(9)&  Fit    & 2.161  & Fit     &2.2246
\\ \hline
$r_d$ (fm)         & 1.971(5)  & 1.979   & 1.987  & 1.984   &1.975
\\ \hline 
$\mu_d$ (n.m.)     & 0.857406(1)& 0.848   & 0.853  & 0.847   &0.847
\\ \hline
$Q_d$ (fm$^2$)     & 0.2859(3)  & 0.280   & 0.276  & 0.291   &0.280
\\ \hline \hline  
\end{tabular}
\end{center}
\caption{Deuteron static properties at NNLO ($r_d$ and $Q_d$) and NLO
($\mu_d$) for a range of deuteron wave functions. At NNLO $\mu_d$
can be reproduced exactly.}
\label{table-compstatic}
\end{table}

On the other hand, it is apparent that $Q_d$ is much more sensitive to
short-distance physics than either $r_d$ or $\mu_d$.  Its value varies
by about 5\% between models with the same pion-range, but different
short-distance, physics.  The counterterm that would absorb this
sensitivity is nominally of $O(eP^5)$, which we estimate to be almost
ten times smaller than is necessary to absorb the variation seen in
Table~\ref{table-compstatic}.  Whether this counterterm should be
promoted to a lower order---as has been argued in
Refs.~\cite{Ch99,PC99}---cannot be properly determined until
higher-order calculations of $Q_d$ are performed and a systematic
study of its renormalization-group evolution is made.

\section{Conclusion}

\label{sec-conclusion}

Chiral perturbation theory, applied to the deuteron four-current in
the fashion suggested by Weinberg~\cite{We90,We91,We92}, produces an
expansion in increasing powers of small momenta ($P$) for the deuteron form
factors $G_C$, $G_Q$, and $G_M$. However, this expansion fails to
reproduce the experimental data at momentum transfers $Q \sim 300$
MeV~\cite{WM01}. The failure, however, lies not in $\chi$PT's description of
{\it deuteron structure}, but with its difficulties in describing
{\it isoscalar nucleon structure}. Applying a chiral expansion to the
ratio of deuteron and nucleon form factors yields NNLO results for 
$G_C$ and $G_Q$ that agree with data to $Q \sim 700$ MeV. $G_C$ and $G_Q$ are 
also relatively insensitive to short-distance physics over this range.

The magnetic form factor, $G_M$, was computed up to NLO, and turns out
to be more sensitive to short-distance physics. This result is anticipated
within the effective theory, since short-distance two-body currents
are suppressed by three powers of $P$ relative to leading
in $G_M$, but are down by two additional powers of $P$ in $G_C$ and $G_Q$.

Deuteron static properties are also well reproduced, although $Q_d$
shows significant variability when different assumptions about
deuteron short-distance physics are made. This may be associated with
the ``$Q_d$-puzzle'': the inability of modern potential models to
reproduce the experimental value for this quantity~\cite{CS98}.  Any
possible resolution of this puzzle within the nuclear effective theory
will require the computation of higher-order effects in $J_0$,
including two-pion-exchange contributions to the deuteron
four-current.

\section*{Acknowledgements}
{I gratefully acknowledge the hospitality of the Institute for Nuclear
Theory, where part of this work was performed. I thank Jiri Adam,
Tae-Sun Park, Martin Savage, and Steve Wallace for useful
conversations on the subjects discussed here. Thanks also to Jacques
Ball, Michel Garcon, and Ingo Sick for providing the parameterizations
of the electron-deuteron scattering data, and to Charlotte Elster,
Evgeni Epelbaum, Ruprecht Machleidt, and Vincent Stoks for supplying
me with deuteron wave functions. This work was supported by the
U.~S. Department of Energy under grants DE-FG03-97ER41014,
DE-FG02-93ER40756 and DE-FG02-02ER41218.}


\begin{thebibliography}{99}
\bibitem{deFW66}
T. deForest and J.~D. Walecka, Adv. Phys. {\bf 15},  1  (1966).

\bibitem{HS81} C.~J.~Horowitz and B.~D.~Serot,
Nucl.\ Phys.\ A {\bf 368}, 503 (1981).

\bibitem{Pi02}
S.~C.~Pieper, K.~Varga and R.~B.~Wiringa,
Phys.\ Rev.\ C {\bf 66}, 044310 (2002)
[arXiv:nucl-th/0206061].

\bibitem{WS98}
R.~B.~Wiringa and R.~Schiavilla,
Phys.\ Rev.\ Lett.\  {\bf 81}, 4317 (1998)
[arXiv:nucl-th/9807037].

\bibitem{CS98}
J. Carlson and R. Schiavilla, Rev. Mod. Phys. {\bf 70},  743  (1998).

\bibitem{BC79}
D.~M. Bishop and L.~M. Cheung, Phys. Rev. C {\bf 20},  381  (1979).

\bibitem{Li65}
I. Lindgren in {\it Alpha, Beta, and Gamma-Ray Spectroscopy}, Vol. 2,
ed. K. Siegbahn (North Holland, Amsterdam, 1965).


\bibitem{Sc84}
M.~E. Schulze {\it et~al.}, Phys. Rev. Lett. {\bf 52},  597  (1984).

\bibitem{Ga94}
M. Garcon {\it et~al.}, Phys. Rev. {\bf C49},  2516  (1994).

\bibitem{Dm85}
V.~F. Dmitriev {\it et~al.}, Phys. Lett. B. {\bf 157},  143  (1985).

\bibitem{Gi90}
R. Gilman {\it et~al.}, Phys. Rev. Lett. {\bf 65},  1733  (1990).

\bibitem{Bo91}
B. Boden {\it et~al.} Z. Phys. {\bf C49}, 175 (1991).

\bibitem{Fe96}
M. Ferro-Luzzi {\it et~al.}, Phys. Rev. Lett. {\bf 77},  2630  (1996).

\bibitem{Bo99}
M. Bouwhuis {\it et~al.}, Phys. Rev. Lett. {\bf 82},  3755  (1999).

\bibitem{Ab00A}
D. Abbott {\it et~al.}, Phys. Rev. Lett. {\bf 84},  5053  (2000).

\bibitem{Ab00B}
D. Abbott {\it et~al.}, Eur. Phys. J. {\bf A7},  421  (2000).


\bibitem{Si01} I.~Sick,
Prog.\ Part.\ Nucl.\ Phys.\  {\bf 47}, 245 (2001)
[arXiv:nucl-ex/0208009].

\bibitem{Wi95}
R.~B. Wiringa, V.~G.~J. Stoks, and R. Schiavilla, Phys. Rev. C {\bf 51},  38
  (1995).

\bibitem{Ar99}
H. Arenh\"ovel, F. Ritz, and T. Wilbois, Phys. Rev. {\bf C61},  034002  (2000).

\bibitem{SP02}
R. Schiavilla and V.~R. Pandharipande, Phys. Rev. {\bf C65},  064009  (2002).


\bibitem{vOG01} 
M.~Garcon and J.~W.~Van Orden,
Adv.\ Nucl.\ Phys,\ {\bf 26}, 293 (2001).
[arXiv:nucl-th/0102049].

\bibitem{GG01}
R.~Gilman and F.~Gross,
J.\ Phys.\ G {\bf 28}, R37 (2002)
[arXiv:nucl-th/0111015].

\bibitem{Be00}
S.~R. Beane {\it et~al.},  in {\em At the frontier of particle physics, vol.
  1}, edited by M. Shifman (World Scientific, Singapore, 2000).

\bibitem{BvK02}
P.~F.~Bedaque and U.~van Kolck,
arXiv:nucl-th/0203055.

\bibitem{We90}
S. Weinberg, Phys. Lett. B. {\bf 251},  288  (1990).

\bibitem{We91}
S. Weinberg, Nucl. Phys. {\bf B363},  3  (1991).

\bibitem{We92}
S. Weinberg, Phys. Lett. B. {\bf 295},  114  (1992).

\bibitem{Be95}
S.~R. Beane, C.~Y. Lee, and U. van Kolck, Phys. Rev. C {\bf 52},  2914  (1995).

\bibitem{WM01}
M. Walzl and U.~G. Mei\ss ner, Phys. Lett. {\bf B513},  37  (2001).

\bibitem{Br95}
V. Bernard, N. Kaiser, and U.-G. Mei\ss ner, Int. Jour. of Mod. Phys. E {\bf 4},
  193  (1995).

\bibitem{AA96}
J. Adam and H. Arenh\"ovel, Nucl. Phys. {\bf A614},  289  (1997).

\bibitem{Sc99}
M.~J. Savage, K.~A. Scaldeferri, and M.~B. Wise, Nucl. Phys. {\bf A652},  273
  (1999).

\bibitem{Ch99}
J.-W. Chen, G. Rupak, and M. Savage, Nucl. Phys. {\bf A653},  386  (1999).

\bibitem{factore}
This peculiarity occurs because we pull out a factor of
$e$ when defining the operators ${\cal O}_\mu$. If we counted $e \sim P$ then
the counting for vertices involving photons would be exactly as for
pion-nucleon interactions.

\bibitem{Ri84}
D.~O. Riska, Prog. Part. Nucl. Phys. {\bf 11},  199  (1984).

\bibitem{unitequivnote} In terms of the notation of Ref.~\cite{Ad93} the
result (\ref{eq:J02B}) corresponds to $\tilde{\mu}=-1$. This occurs
because the field-theoretic manipulations used to arrive at
Eq.~(\ref{eq:J02B}) assume that the fields represent physical
particles, i.~e. they are on-shell. This choice has been the standard
one for computing $\chi$PT kernels for interactions with light nuclei,
see, e.g.~\cite{Be97,Be99}.

\bibitem{Pa95}
T.-S. Park, D.-P. Min, and M. Rho, Nucl. Phys. {\bf A596},  515  (1996).

\bibitem{Pa00}
T.-S. Park, K. Kubodera, D.-P. Min, and M. Rho, Phys. Lett. {\bf B472},  232
  (2000).

\bibitem{Ep99}
E. Epelbaum, W. Gl\"ockle, and U.-G. Mei\ss ner, Nucl. Phys. {\bf A671},  295
  (2000).

\bibitem{EM01}
D.~R. Entem and R. Machleidt, Phys. Lett. {\bf B524},  93  (2002).

\bibitem{PC99}
D.~R. Phillips and T.~D. Cohen, Nucl. Phys. {\bf A668},  45  (2000).

\bibitem{St93}
V.~G.~J. Stoks, R.~A.~M. Klomp, M.~C.~M. Rentmeester, and J.~J. de~Swart, Phys.
  Rev. C {\bf 48},  792  (1993).

\bibitem{Fr80}
J.~L. Friar, Phys. Rev. C {\bf 22},  796  (1980).

\bibitem{Ad93}
J. Adam, H. Goller, and H. Arenh\"ovel, Phys. Rev. C {\bf 48},  470  (1993).

\bibitem{Fo99}
J.~L.~Forest,
Phys.\ Rev.\ C {\bf 61}, 034007 (2000)
[arXiv:nucl-th/9905063].



\bibitem{Br92}
V.~Bernard, N.~Kaiser, J.~Kambor and U.~G.~Mei\ss ner,
Nucl.\ Phys.\ B {\bf 388}, 315 (1992).

\bibitem{KM99}
B.~Kubis and U.~G.~Mei\ss ner,
Nucl.\ Phys.\ A {\bf 671}, 332 (2000)
[Erratum-ibid.\ A {\bf 692}, 647 (2001)]
[arXiv:hep-ph/9908261].

\bibitem{He98}
V.~Bernard, H.~W.~Fearing, T.~R.~Hemmert and U.~G.~Mei\ss ner,
Nucl.\ Phys.\ A {\bf 635}, 121 (1998)
[Erratum-ibid.\ A {\bf 642}, 563 (1998)]
[arXiv:hep-ph/9801297].

\bibitem{Me95}
P.~Mergell, U.~G.~Mei\ss ner and D.~Drechsel,
Nucl.\ Phys.\ A {\bf 596}, 367 (1996)
[arXiv:hep-ph/9506375].

\bibitem{Si81}
G.~G. Simon and C.~Schmitt and V.~H.~Walther, Nucl. Phys. {\bf A364},
285 (1981).

\bibitem{Au85} 
S. Auffret {\it et~al.}, Phys. Rev. Lett. {\bf 54},  649 (1985).

\bibitem{Cr85}
R. Cramer {\it et al.}, Z. Phys. {\bf C29}, 513 (1985).

\bibitem{SS01}
R.~Schiavilla and I.~Sick,
Phys.\ Rev.\ C {\bf 64}, 041002 (2001)
[arXiv:nucl-ex/0107004].

\bibitem{Be97}
S.~R. Beane {\it et~al.}, Nucl. Phys. {\bf A618},  381  (1997).

\bibitem{Be99}
S.~R. Beane, M. Malheiro, D.~R. Phillips, and U. van Kolck, Nucl. Phys. {\bf
  A656},  367  (1999).


\end{thebibliography}

\end{document}